\documentclass[preprints,article,accept,moreauthors,pdftex]{Definitions/mdpi}

\firstpage{1} 
\pubvolume{xx}
\issuenum{x}
\articlenumber{x}
\pubyear{x}
\copyrightyear{x}
\history{~}

\Title{Stirred Kardar-Parisi-Zhang equation with quenched random noise: Emergence of induced nonlinearity}

\Author{P.~I.~Kakin,$^{1}$\ M.~A.~Reiter,$^{1}$\ M.~M.~Tumakova,$^{2}$\
N.~M.~Gulitskiy$^{1}$\ and\ N.~V.~Antonov$^{1,3}$}

\address{%
	$^{1}$ \quad Department of Physics,  
	Saint Petersburg State University, Universitetskaya nab.~7/9, St. Petersburg, 199034, Russia;\\
		$^{2}$ \quad  L.D.~Landau Institute for Theoretical Physics,  Ak.~Semenova~1-A, Chernogolovka, 142432, Moscow Region, Russia; \\
		$^3$  \quad 
	 N.N. Bogoliubov Laboratory of Theoretical Physics, Joint Institute for Nuclear Research, Dubna, 141980, Moscow Region, Russia.
	}

\abstract{We study the stochastic Kardar-Parisi-Zhang equation for kinetic roughening 
where the time-independent (columnar or spatially quenched) Gaussian random noise 
$f(t,{\bf x})$ is specified by the pair correlation function 
$\langle f(t,{\bf x})f(t',{\bf x'}) \rangle \propto \delta^{(d)} ({\bf x-x'})$, 
$d$ being the dimension of space. 
The field-theoretic renormalization group analysis shows that the effect of
turbulent motion of the environment (modelled by the coupling with the velocity field described by the Kazantsev–Kraichnan statistical ensemble for an incompressible fluid)
gives rise to a new nonlinear term, quadratic in the velocity field. 
It turns out that this ``induced'' nonlinearity strongly affects the scaling
behaviour in several universality classes (types of long-time, large-scale asymptotic regimes) 
even when the turbulent advection appears irrelevant in itself. 
Practical calculation of the critical exponents (that determine the universality classes) is performed to the first order of the double expansion in $\varepsilon=4-d$ and the 
velocity exponent $\xi$  (one-loop approximation). 
As is the case with most ``descendants'' of the Kardar-Parisi-Zhang model, 
some relevant fixed points of the renormalization group equations 
lie in ``forbidden zones,''  i.e.  in those corresponding to 
negative kinetic coefficients or complex couplings. This persistent phenomenon 
 in stochastic non-equilibrium models 
requires careful and inventive physical interpretation.
}


\begin{document}

	
\section{Introduction \label{Intro}}


The Kardar-Parisi-Zhang (KPZ) model was proposed in \cite{KPZ} to describe evolution of an interface that separates a randomly growing substance from the rest of the system. 
 As the interface evolves due to intrinsic dynamics and external disturbances, it becomes progressively ``rough.'' This joint effect of various deterministic or/and random entries results in what is known as kinetic roughening~\cite{rost1}. Flame fronts, surfaces of tumours or bacterial colonies, and landscape profiles are all examples of interfaces undergoing kinetic roughening; see, e.g. \cite{rost2} and references therein.\footnote{
  To be precise, an equivalent model was introduced much earlier in terms of a vector field in a seminal paper by Forster, Nelson and Stephen \cite{FNS1}. There, among other relevant models,
  the stochastic  $d$-dimensional generalization of the Burgers equation was studied in 
  connection with problem of long-time tails in hydrodynamic description of  fluids.}
  
The KPZ equation is one of  the simplest semi-phenomenological models of kinetic roughening. 
The model itself is a nonlinear stochastic differential equation for a smoothed height profile~$h(t,{\bf x})$ of the moving interface. It assumes that the growth is lateral and that its rate is a smooth function of the height gradient.
Then, the  leading term in the gradient expansion determines the nonlinearity of the KPZ model. Another term, linear in $h$, incorporates the ``surface tension'' or the forces of any kind that make the interface smoother. A random noise 
mimics various microscopic degrees of freedom that can influence the roughening dynamics of the interface.

From physics viewpoints, kinetic roughening is a representative example of non-equilibrium phenomena that occur in a wide variety of complex physical systems evolving due to intrinsic dynamics and undergoing extra 
disturbances.\footnote{Giorgio Parisi was awarded the Nobel Prize in Physics 2021 ``for the discovery of the interplay of disorder and fluctuations in physical systems from atomic to planetary scales'' \cite{Nobel}.}
It was argued that it is such competition that gives rise to nontrivial patterns \cite{Mitch}.
Indeed, many of non-equilibrium  systems can evolve to critical states without the fine tuning of control parameters (in contrast to near-equilibrium critical systems).
Examples are provided by dissipative driven open systems \cite{Zia}, turbulence \cite{UFN,RedBook}
and, in a more general context, by systems revealing the so-called self-organized criticality, phenomenon  observed in numerous physical, biological, chemical, neural and social systems, etc. \cite{Bak,Bak3,Col1,Col2}.

As such systems exhibit universal scaling behaviour, fluctuations of growing surfaces can be viewed as the most pictorial representative for 
a wide range of phenomena with the same critical behaviour (universality class).
Thus, the KPZ model, with its simplest make-up, sometimes is referred to as a kind of non-equilibrium analog of the Ising model in equilibrium phase transitions.

That is why a significant number of papers concerned with the KPZ model are published every year as the KPZ universality class is discovered in a new system  or new features of the KPZ model are explored 
\cite{freshrev,ff,Sinc,Uppergt41,UpperInf,Polaritons,Polaritons1,Polaritons2,CanetColor}. For example, in the recent paper \cite{Sky}, the KPZ universality class was established in the morphology of the modern urban skyline measured in cities throughout the Netherlands.

However, the standard field-theoretic perturbative renormalization group (RG) analysis does not reveal an infrared (IR) attractive fixed point of the RG equations for the KPZ model~\cite{11,111} in the physical area of parameters. If the 
appropriate fixed point does nevertheless exist, it cannot be accessible within any kind of perturbative treatment.
This point would correspond to the rough phase or, to be precise, to the non-trivial asymptotic behaviour of the interface in the IR range (which implies that times and distances are large in comparison with the characteristic microscopic scales), i.e.  to kinetic roughening (or critical scaling).  

The functional RG is probably the only existing approach that gives access to that fixed point~\cite{Canet,Canet1,Canet2,Canet4} making it ``essentially non-perturbative.'' 

Other open questions include, e.g. the value of the upper critical dimension~\cite{Uppergt41, UpperInf}, \cite{LK,LK1,UpperInf0,Upper4,Uppergt4} or the random noise interpretation~\cite{Howard,Cooper}.

All of these facts suggest that instead of a more sophisticated analysis, the KPZ model may need modifications or adjustments that might lead to a drastic change in the RG analysis. It does seem as if the KPZ model may be sensitive to various extensions and modifications; e.g., the simple extension  turns the model into one with infinite number of coupling constants~\cite{Pav,Pav2}. 

One of the possible modifications consists of choosing a time-independent (spatially quenched or columnar) noise with the correlation function
\begin{equation}
\langle f(x)\,f(x') \rangle = D_{0}\,
\delta^{(d)}({\bf x}-{\bf x}'), \quad D_0>0
\label{forceStat}
\end{equation}
instead of the white in-time random noise (that differs from~(\ref{forceStat}) by additional Dirac's function $ \delta(t-t')$) that was used in the original KPZ model.
Here $x=(t,{\bf x})$ are the space-time coordinates, the brackets $\langle\dots \rangle$ stand for appropriate averaging, and $d$ is the dimension of space. 

The spatially quenched noise~(\ref{forceStat}) was suggested in~\cite{Caldarelli} to model landscape erosion where non-erodible (``quenched'') regions may be the main reason behind the scaling~\cite{Czi}. 

The KPZ model is usually considered with a more general form of a quenched disorder that includes dependence on the height of the profile $\Delta(h-h')$ \cite{JP,KimKim,KimKimKim}. This disorder is often used in the study of driven interfaces in random media \cite{Hinri} (also confer~\cite{Hinri} for a detailed review of the types of quenched noises). However, the factor $\Delta(h-h')$ is a hard obstacle for analytical approaches \cite{Nara}.

Besides its relative simplicity, the noise~(\ref{forceStat}) also stands aside for its connection with nonuniversality (see \cite{Janssen,Moreira,Webman} in relation to directed percolation, \cite{Delamotte} in relation to erosion of landscapes,  and \cite{Vitalik} in relation to self-organised criticality).

In this paper we propose to analyse the version of the KPZ equation where the white noise is replaced with the time-independent spatially quenched noise~(\ref{forceStat}). However, one could include the spatially quenched noise in another way, namely, by coupling the conserved KPZ equation (i.e. the modification of the model with conservation law) with the spatially quenched noise \cite{CKPZSQN}. Note that the resulting model is vastly different.
 
Another possible modification of the KPZ model involves inclusion of the motion of the environment. Critical behaviour of nearly-equilibrium nearly-critical systems can be dramatically affected by the motion of the medium either disappearing altogether or acquiring new unexpected features~\cite{Onuki2,Satten2,Nelson,AHH,Alexa,AIK,AKM}. Considering that environment motion is almost impossible to exclude in real experimental settings, it is important to account for it while studying critical behaviour. 

Recent attempts \cite{Us,AKL,Script} revealed that turbulent or random environment (modelled either by stochastic Navier-Stokes equation or by a ``rapid-change'' Kazantsev-Kraichnan velocity ensemble) dominates scaling and ``washes away'' the kinetic roughening. 

In this paper, we apply the field theoretic RG 
to a nonlinear non-equilibrium nearly-critical system, subjected to a quenched disorder and turbulent environment. The system is described by the KPZ equation, while the disorder is described by the Gaussian time-independent spatially quenched noise~(\ref{forceStat}). 
The environment is modelled by the ``synthetic'' Gaussian ensemble with vanishing correlation time known as the Kazantsev-Kraichnan ensemble \cite{FGV}.

We discover that coupling with the turbulent velocity field leads to an emergence of a new nonlinearity that must be included in the model to make it renormalizable. RG analysis shows that there are six regimes of critical behaviour; critical exponents are calculated for every regime
in the leading order of the double expansion in $\varepsilon=4-d$ and velocity exponent $\xi$ (one-loop approximation). The most realistic values of parameters ($d=1,2,3$ and $\xi=4/3$) correspond to the regime where the turbulent advection is irrelevant (in the sense of Wilson) while the new nonlinearity is relevant along with the KPZ nonlinearity.

The plan of the paper is as follows: the problem is described in Sect. 2; Sect. 3 details renormalization procedure up to calculation of renormalization constants; fixed points of the RG equation, their stability regions, and corresponding critical exponents are considered in Sect. 4; Sect. 5 contains conclusion and discussion of the implications.
	

\section{Formulation of the problem}


Kinetic roughening of growing interfaces can be described by a power law for asymptotic behaviour of the so-called structure functions in the IR range. It could be entered as following: 
\begin{equation}
S_{n}(t,r)=
\langle\left[h(t,{\bf x}) - h(0,{\bf 0})\right]^{n}\rangle \simeq
r^{n\chi}\, F_{n} (t/r^{z}), \quad  r=|{\bf x}|.
\label{scaling}
\end{equation}
Here $h(t,{\bf x})$ stands for the height of the surface profile (here and below, $t$ and ${\bf x}$ are the time and the space coordinates, respectively), while the averaging~$\langle\dots\rangle$ is performed over the statistical ensemble. The roughness exponent $\chi$, the dynamical exponent $z$ and the universal scaling functions $F_{n}(\cdot)$ determine the universality class of the scaling behaviour.

To calculate the critical exponents $\chi$ and $z$, we perform the RG analysis of the model which consists of the equation for the interface growth and statistical ensemble for the velocity field that models environment motion.

The KPZ model that describes interface growth is a nonlinear differential equation for the field $h(x)=h(\mathbf{x},t)$: 
	\begin{equation}
		\partial_t h=\nu_0\partial^2 h+\frac{\lambda_0}{2} (\partial h)^2+f.
		\label{eq1}
	\end{equation}
Here $\partial_t=\partial / \partial_t$, $\partial_i=\partial / \partial {x_i}$, $\partial^2 =\partial_i \partial_i$, $(\partial h)^2 =(\partial_i h) (\partial_i h)$, $i=1,...,d$ and $d$ is the dimension of the space. Summation over repeated tensor indices is implied throughout the paper. The parameter $\nu_0 > 0$ corresponds to the ``surface tension,'' and $\lambda_0$ (can be either positive or negative) stays by the nonlinear term of the equation. The nonlinearity models lateral growth or decay. Let us first set $\lambda_0=1$ as a non-trivial $\lambda_0$ can always be scaled out. 
	
The random noise $f$ is supposed to simulate the processes occurring at small scales (which means the smallness of the noise correlation radius in comparison with the distances we are interested in). So it is reasonable to choose the noise correlation function in the form of a spatial $\delta$ function. A $\delta$ function in time would correspond to a vanishing correlation time which is also a reasonable assumption. Here, however, we choose the spatially quenched noise~(\ref{forceStat}) discussed in Sec. \ref{Intro}.

This choice violates the Galilean symmetry of the original deterministic equation (the symmetry is preserved for the white
in-time noise with $\langle f(x)f(x') \rangle \propto \delta(t-t')$).

Let us proceed with the description of the turbulent motion of the environment (e.g. some fluid). The velocity of the mixing field is represented by the Kazantsev–Kraichnan statistical Gaussian ensemble (see, e.g. \cite{FGV}), i.e. the distribution with zero mean and the correlation function of the form
\begin{equation}
	\begin{gathered}
		\langle v_i(t,\mathbf{x}) v_j(t',\mathbf{x'})\rangle=\delta(t-t') D_{ij}(\mathbf{x}-\mathbf{x'}), \\
		D_{ij}(\mathbf{r})=\frac{B_0}{(2 \pi)^d} \int_{k>m} \frac{dk}{ k^{d+\xi}}P_{ij}(\mathbf{k}) e^{i({\bf k \cdot r})}.
	\end{gathered}
\label{eq3}
\end{equation}
Here $P_{ij}(\mathbf{k})=\delta_{ij}-k_i k_j/k^2$ is the transverse projector, it reflects the incompressibility of the fluid ($\partial_i v_i = 0$); $k=|\mathbf{k}|$ is the wave number, $B_0>0$ is a positive amplitude. The cutoff $k>m$ serves as an IR regularization.
The advection by the velocity field is provided by the ``minimal'' replacement $\nabla_t h~=~\partial_t h~+~(v_i \partial_i) h$ in Eq.~(\ref{eq1}). 


\section{Renormalization of the model}
	

\subsection{Field theoretic formulation of the model}


According to the Dominicis-Janssen theorem the stochastic problem~(\ref{forceStat}),~(\ref{eq1}),~(\ref{eq3}) could be reformulated \cite{Vasiliev} as the field theoretic one with an increased number of fields $\Phi=\{h,h',v_i\}$. The equivalence between them implies the equivalence between the correlation functions of the problem~(\ref{forceStat}),~(\ref{eq1}),~(\ref{eq3}) and the Green's functions of the field theory with the action functional $S(\Phi)=S_h(\Phi)+S_v(\Phi)$, where
	\begin{equation}
		\begin{gathered}
		S_h(\Phi)=\frac{1}{2} h' D_0 h'+h' \{ -\nabla_t h +\nu_0 \partial^2 h+\frac{1}{2}(\partial h)^2\}, \\
		S_v(\Phi)=-\frac{1}{2} \int dt \int d {\bf x} \int d {\bf{x'}} v_i(t, {\bf{x}}) D^{-1}_{ij}({\bf{x}}-{\bf{x'}}) v_j(t,\bf{x'}).
		\end{gathered}
	\label{eq4}
	\end{equation}
Here $D^{-1}_{ij}(\bf x-\bf x')$ is the kernel of the operator $D^{-1}_{ij}$ which is inverse to~(\ref{eq3}).
The integrations over the appropriate arguments $t \text{, } {\bf x}$ are implied for all terms in the expressions for the action functionals, e.g.,
	\begin{equation}
		\begin{gathered}
	\frac{1}{2}\, h'D_0\,h' = \frac{1}{2}\,D_0\int dt \int dt' \int d{\bf x}\, h'(t',{\bf x})h'(t,{\bf x}),\\
h'\nabla_t h = \int dt \int d{\bf x}\,h'(t,{\bf x})\nabla_t h(t,{\bf x}).
\end{gathered}
\end{equation}
The parameters ${\tilde g_0}, {\tilde w_0}$ serve as the coupling constants (``charges''):
\begin{equation}
{\tilde g_0}=D_0/{\nu_0^4} \sim \Lambda^\varepsilon, \quad {\tilde w_0}=B_0/{\nu_0} \sim \Lambda^\xi.
\end{equation}
The last relations follow from the dimensional analysis (detailed in section \ref{UV}); $\Lambda$ defines the typical ultraviolet momentum scale.

The expressions for the bare propagators are obtained by considering the terms quadratic in the fields fields in the action functional~(\ref{eq4}). The propagators have the following form in the frequency--momentum representation:
\begin{equation}
	\begin{gathered}
		\langle h' h' \rangle_0 = 0\text{, } \quad
		\langle h h' \rangle_0=\langle h' h \rangle_0^{*}=\frac{1}{-i \omega + \nu_0 k^2} \\
		\langle h h \rangle_0=\frac{2 \pi  D_0 \delta(\omega)}{\nu_0^2 k^4} \text{, }\quad
		\langle v_i v_j \rangle_0=\frac{B_0}{k^{d+\xi}}P_{ij}(k).
	\end{gathered}
	\label{propag}
\end{equation}
The propagators~(\ref{propag}) include the amplitudes $B_0$, $D_0$. This is not ideal because while we can still proceed with the calculations, critical dimensions for the fixed points of RG equations with coordinates equal to zero will have to be readjusted. 
Indeed, the propagators that involve bare charges should be treated carefully in the vicinity of trivial renormalized charges, e.g. when a fixed point has coordinates ${\tilde g^*}=0$ or ${\tilde w^*}=0$. Otherwise, the expressions for the critical dimensions naively derived for such points using the standard formulas~(\ref{eq20}) can be wrong. 
To avoid that complication, let us dilate the fields $\{h,h',v_i\}$ so that the couplings  ${\tilde g_0}, {\tilde w_0}$ are removed from quadratic terms.

The appropriate choice for the rescaling is $h'D_0h'\rightarrow h'h'$ and $v_i v_j/B_0\rightarrow v_i v_j$, i.e. we exchange the set of fields $\{h,h',v_i\}$ with the set $\{h\,{\tilde g_0}^{-1/2}\nu_0^{-2},\, h'\,{\tilde g_0}^{1/2}\nu_0^{2},\, v_i\,{\tilde w_0}^{-1/2}\nu_0^{-1/2}\}$. Let us also pass to new charges $g_0^2=\tilde g_0$, $w_0^2=\tilde w_0$.

We arrive at the following action functional:
\begin{equation}
		\begin{gathered}
		S(\Phi)=\frac{1}{2} h' h'+h' \{ -\partial_t h - w_0 \nu_0^{\frac{1}{2}} (v_i\partial_i)h +\nu_0 \partial^2 h+\frac{1}{2}g_0\nu_0^2(\partial h)^2\} + S_v(\Phi),\\
		S_v(\Phi)=-\frac{1}{2} \int dt \int d {\bf x} \int d {\bf{x'}} v_i(t, {\bf{x}}) D^{-1}_{ij}({\bf{x}}-{\bf{x'}}) v_j(t,\bf{x'}),\\
		D_{ij}({\bf{r}})=\frac{1}{(2 \pi)^d}\int_{k>m} \frac{dk}{ k^{d+\xi}}P_{ij}({\bf{k}}) e^{i({\bf k \cdot r})}. 
		\end{gathered}
	\label{eq41}
	\end{equation}
Note that we did not change notations for the fields and for the operator $D_{ij}({\bf{r}})$ even though those quantities were rescaled.	


\subsection {UV divergences and renormalization \label{UV}}


To eliminate UV divergences, the renormalization procedure is applied.
Analysis of UV divergences is based on canonical dimensions, see, e.g. \cite{Vasiliev}.
Dynamic models have two independent scales: a time scale $[T]$ and a space scale $[L]$, therefore the canonical dimension of any quantity $F$ is determined by two numbers, namely, by the 
frequency dimension $d_F^\omega$ and by the momentum dimension $d_F^k$: 
\begin{equation}
\left[F\right] \sim \left[T\right]^{-d_F^\omega} \left[L\right]^{-d_F^k}.
\end{equation}
Values of canonical dimensions are found from the requirement that all terms of an action functional 
be dimensionless with respect to both canonical dimensions; the obvious normalization conditions are
\begin{equation}
	d_{k_i}^k = -d_{x_i}^k = 1, \, d_{k_i}^\omega = d_{x_i}^\omega = 0, \, d_\omega^k = d_t^k = 0, \, d_{\omega}^{\omega}=-d_t^{\omega}=1.
\end{equation} 

A total canonical dimension $d_{F}$ is defined by the expression 
$d_{F} = d_{F}^{k} + 2d_{F}^{\omega}$. The factor $2$ follows from the fact that $\partial_t \propto {\bf \partial}^2$ in the free theory. All the canonical dimensions for the theory~(\ref{eq41}) are presented in the Table~\ref{tab1}. Parameters $x_0$, $x$, $\mu$ will be defined later on. 

The model is logarithmic (all coupling constants become dimensionless) at $\varepsilon=0$, i.e.  at $d=4$, and $\xi=0$.

\begin{table}[H] 
\caption{Canonical dimensions of the fields and the parameters in the theory~(\ref{eq41}); $\varepsilon=4-d$. \label{tab1}}
\begin{center}
\begin{tabular}{ |c|c|c|c|c|c|c|c|c|c| }
		\hline
		F&$h$&$h'$ &$v_i$&$\nu_0,\nu$&$g_0$&$w_0$&$x_0$&$\mu$& $g$,$w$,$x$ \\ \hline
		$d^\omega_F$&$-1$&$1$&$1/2$&1&0&0&0&0&0 \\ \hline
		$d^k_F$&$d/2$&$d/2$&$-\xi/2$&$-2$&$\varepsilon/2$&$\xi/2$&$\xi-\varepsilon/2$&1&0 \\ \hline
		$d_F$&$d/2-2$&$d/2+2$&1$-\xi/2$&0&$\varepsilon/2$&$\xi/2$&$\xi-\varepsilon/2$&1&0 \\
		\hline
	\end{tabular}
	\end{center}
	\end{table}

A total dimension $d_\Gamma$ of a 1-irreducible Green's function $\Gamma$ that involves $N_h$ fields $h$, $N_{h'}$ fields $h'$, and $N_v$ fields $v_i$ is determined by the following expression 
\begin{equation}
d_\Gamma=d+2-d_h N_h-d_{h'} N_h' -d_{v_i} N_v.
\end{equation}
In the logarithmic theory, $d_\Gamma$ coincides with the formal index of UV divergence $\delta$ of the corresponding
Green's function. Thus, the divergent part of the function $\Gamma$ and the possible counterterms are polynomials of degree $\delta=d_\Gamma|_{d=4}$. 

In the case of the theory~(\ref{eq41}), the real divergence index differs from the formal one: $\delta'=\delta-N_h$. This is due to the fact that the field $h$ enters the action functional only in the form of a spatial derivative. 

Considering the last condition, one can list possible counterterms: $\partial^2 h'$, $\partial_t h'$ (these two are responsible for renormalization of the mean value of $h$), $h'h'$, $h' \partial^2 h$, $h' \partial_t h$, $h'(\partial h)^2$, $h'(v_i \partial_i)h$ (these five already exist in the action functional), $h'(\partial_i v_i)$ (this one vanishes owing to the incompressibility of the fluid), and, finally, $h'{\bf v}^2$ which is a new counterterm. 

For the model to be renormalizable,  the new term $\frac{x_0}{2\nu_0}h'{\bf v}^2$ must be added to the action functional~(\ref{eq41}):
\begin{equation}
		\begin{gathered}
		S(\Phi)=\frac{1}{2} h' h'+h' \left\{ -\partial_t h - w_0 \nu_0^{\frac{1}{2}} (v_i\partial_i)h +\nu_0 \partial^2 h+\frac{1}{2}g_0\nu_0^2(\partial h)^2 + \frac{1}{2}\frac{x_0}{\nu_0}{\bf v}^2 \right\} + S_v(\Phi).
		\end{gathered}
	\label{eq42}
	\end{equation}
The coupling constant corresponding to the new nonlinearity is notated as $x_0$. The factor $\nu_0^{-1}$ appears from the dimensionality considerations.

Now that we ensured multiplicative renormalizability of the theory~(\ref{eq42}), bare fields and parameters can be expressed in terms of their renormalized counterparts:
\begin{equation}
	\begin{gathered}
	h_0=Z_h h, \quad h'_0=Z_h' h', \quad {\bf v}_0=Z_v {\bf v}, \quad \\
	g_0=Z_g \mu^{\varepsilon/2} g, \quad w_0=Z_w \mu^{\xi/2} w, \quad x_0=Z_x \mu^{\xi-\varepsilon/2} x. 
	\end{gathered}
\label{eq8}
\end{equation} 
Here we added the subscript $0$ to the fields to differentiate them from the renormalized ones. The renormalization mass $\mu$ is an additional parameter of the renormalized theory while the set $\{Z\}$ are renormalization constants. 

The renormalized action has the form
\begin{equation}
	\begin{gathered}
	S_R(\Phi) = \frac{1}{2} Z_1 h' h'+h' \left\{ \partial_t h-Z_2 {w} \nu^2 (v_i \partial_i) h+Z_3 \nu \partial^2 h+\frac{1}{2}Z_4{g}\nu^2(\partial h)^2 +\right. \\ \left.\frac{1}{2} Z_5 \frac{x}{{\nu}} {\bf v}^2 \right\} 
	+ S_v.
	\label{eq9}
		\end{gathered}
\end{equation}
The constants $\{Z\}$ are found from the condition that the corresponding Green's functions be UV  finite in the given order of the perturbation theory.

For the action~(\ref{eq9}), the relations between the renormalization constants are: 
\begin{eqnarray}
	&&Z_h=Z_1^{-1/2}; \quad Z_{h'}=Z_1^{1/2}; \quad  
		Z_\nu=Z_3; \quad Z_v=1;\nonumber\\
		 &&Z_g=Z_1^{1/2} Z_3^{-2} Z_4; \quad 
		Z_w=Z_2 Z_3^{-1/2}; \quad   
		Z_x=Z_1^{-1/2}Z_5 Z_3.
\end{eqnarray}
Using MS scheme and one-loop approximation, we arrive at the following results (see \cite{Us} for a detailed example of one-loop calculations): 

\begin{equation}
	\begin{gathered}
		Z_1=1-\frac{\hat{g}^2}{2 \varepsilon}, \\
		Z_2=1+\frac{\hat{g}^2}{d \varepsilon}-\hat{x}\hat{g}\left(1-\frac{1}{d}\right)\frac{1}{2 \xi}, \\
		Z_3=1+\frac{\hat{g}^2}{d \varepsilon}-\hat{w}^2\left(1-\frac{1}{d}\right)\frac{1}{2 \xi}, \\
		Z_4=1+\frac{\hat{g}^2}{d \varepsilon}-\hat{w}^2\left(1-\frac{1}{d}\right)\frac{1}{2 \xi}, \\
		Z_5=1+\frac{\hat{g}\hat{w}^2}{d\hat{x} \varepsilon}-\hat{x}\hat{g}\left(1-\frac{1}{d}\right)\frac{1}{2 \xi}. 
	\end{gathered}
\label{eq11}
\end{equation}
Here $\hat{f}=f \sqrt{S_d/(2 \pi)^d}$ for any $f=\{g,w,x\}$, where $S_d =2\pi^{d/2}/\Gamma(d/2)$ is the area of the unit  sphere in $d$-dimensional
space.


\section{RG equation, fixed points, critical exponents \label{FPs}}



\subsection{RG equation, RG functions}


The Green's functions of the theory~(\ref{eq41}) can be expressed in terms of their renormalized counterparts in the following way:
\begin{equation}
	G(e_0,\dots)=Z_h^{N_h} Z_{h'}^{N_{h'}} Z_v^{N_v}G_R(\mu, e_,\dots ).
	\label{eq12}
\end{equation}
Here $e_0$ denotes the full set of the bare parameters $\{\nu_0,g_0,w_0,x_0\}$ while $e$ stands for renormalized ones; other arguments (times, momenta etc.) are denoted with the ellipsis. $N_h$, $N_{h'}$, $N_v$ are numbers of the respective fields in the function $G$.

We apply the operator $\mu \partial_\mu$, taken at fixed bare parameters, to both sides of the expression~(\ref{eq12}) to obtain the basic RG equation:
\begin{equation}
	(D_\mu+\beta_g \partial_g+\beta_w \partial_w+\beta_x \partial_x-\gamma_\nu D_\nu -\gamma_{G_R})G_R=0
	\label{eq13}
\end{equation}
Here $D_{\mu}=\mu \partial_{\mu} $ and $D_{\nu}=\nu \partial_{\nu} $; anomalous dimensions $\gamma_{\nu}$, $\gamma_{G_R}$ and beta functions $\beta_g$, $\beta_w$, $\beta_x$ are so-called RG functions and they are defined in the standard way:
\begin{equation}
	\gamma_f=D_{RG} \ln Z_f \text{, }\quad \beta_f=D_{RG} f,
	\label{eq14} 
\end{equation}
where $f=\{g,w,x\}$ and $D_{RG}=\mu \partial_{\mu}|_{e_0}$.
From~(\ref{eq8}),~(\ref{eq14}) the following relations can be obtained:
\begin{eqnarray}
\begin{gathered}
	\beta_g=-g(\varepsilon/2+\gamma_g), \\
	\beta_w=-w(\xi/2+\gamma_w), \\
	\beta_x=-x(\xi-\varepsilon/2+\gamma_x).
\end{gathered}
\label{kb}
\end{eqnarray}
The corresponding anomalous dimensions in one-loop approximation are (hereafter we omit the notation $\hat{ }$ ):
\begin{eqnarray}
    \begin{gathered}
	\gamma_g=g^2/2-3w^2/8, \\
	\gamma_w=3xg/8-g^2/8-3w^2/16,\\
	\gamma_x=-g^2/2-w^2g/4x+3xg/8+3w^2/8.
    \end{gathered}
\end{eqnarray}


\subsection{Fixed points, stability regions}


Critical exponents that characterize IR asymptotic behaviour of a system are associated with IR attractive fixed points of the RG equation derived for the corresponding field theory (see, e.g. \cite{Vasiliev}).
Values of the renormalized charges $(g^*,w^*,x^*)$ serve as coordinates of the fixed points; they are found from the system of equations $\beta_f(g^*,w^*,x^*)=0$ for all $f =\{g,w,x\}$. Matrix $\Omega_{ij}(g,w,x)=\partial \beta_{f_j}/\partial f_i$ (where $f$ again is a set of charges $\{g,w,x\}$) determines the type of fixed point. For IR attractive points, ${\rm Re} (\lambda_k)~>~0 \text{ for } \forall k=\{1,2,3\}$, where $\{\lambda_k \}$ is a complete set of eigenvalues of the matrix $\Omega_{ij}(g^*,w^*,x^*)$. The region of system's parameters where a given point is IR attractive is referred to as an IR stability region of this point.

The system of beta functions~(\ref{kb}) reads
\begin{equation}
\label{betas}
	\begin{gathered}
		\beta_g=-g\left(\varepsilon/2+g^2/2-\frac{3}{8}w^2\right), \\
		\beta_w=-w\left(\xi/2-\frac{g^2}{8}+\frac{3}{8} xg-\frac{3}{16} w^2\right), \\
		\beta_x=-x\left(\xi-\varepsilon/2-g^2/2-\frac{1}{4}w^2g/x+\frac{3}{8} xg+\frac{3}{8} w^2\right).
	\end{gathered}
\end{equation}

The system has the following solutions: 
\begin{itemize}
\item FP1a (fixed point 1a) $g^*=w^*=x^*=0$ 
with IR stability region $\varepsilon<0, \xi<0, \xi~<~\varepsilon/2$. 

\item FP2 $g^{*2}=-\varepsilon$, $x^*=w^*=0$ with IR stability region $\varepsilon<-4\xi, \xi<0, \varepsilon>0$. 
Note that this point is actually two points (as there are two roots of the equation $g^{*2}=-\varepsilon$) that share the same stability region. 
\item FP3 $g^{*2}=-4\xi-2\varepsilon$, $w^{*2}=-16\xi/3-4\varepsilon/3$, $x^*=(-16\xi/3-4\varepsilon/3)/g$ 
with IR stability region $\varepsilon>-4\xi, \xi<0$ (for all 4 combinations of the roots). 

\item FP4 $g^{*2}=4\xi-2\varepsilon$, $w^{*2}=16\xi/3-4\varepsilon/3$, $x=2g/3$ 
with IR stability region $\varepsilon>4\xi, \xi>0$ (for all 4 combinations of the roots). 

\item FP6 $g^{*2}=-\varepsilon$, $w^*=0$, $x^*=-8\xi/3g$ 
with IR stability region $\varepsilon<4\xi, \xi>0, \varepsilon>0$ (for both roots).

\item FP7 $g^*=x^*=0$, $w^{*2}=8\xi/3$
with empty IR stability region (for both roots). 
\end{itemize}

The listed fixed points must be supplemented with the solutions that account for marginal values of the charges. To find them, let us pass to a new set of charges with $\beta-$functions that do not involve terms with nontrivial denominators. The appropriate substitution is a set $\{g, y=xg, \alpha=w^2/y\}$ with  a system of $\beta-$functions:
\begin{equation}
	\begin{gathered}
		\beta_g=-g\left(\varepsilon/2+g^2/2-\frac{3}{8}\alpha y\right), \\
		\beta_y=-y\left(\xi-\frac{\alpha g^2}{4}+\frac{3}{8} y\right), \\
		\beta_{\alpha}=-\alpha\left(-g^2/4-\frac{3}{8}\alpha y+\frac{3}{8} y+\frac{1}{4} \alpha g^2\right).
	\end{gathered}
\end{equation}

Additional fixed points include
\begin{itemize}
\item FP1b $g^{*2}=0, y^*=0$, arbitrary $\alpha^*$ with IR stability region  $\varepsilon < 0, \xi < 0$;
\item FP5 $g^{*2}=0, y^*=-8\xi/3, \alpha^*=0$ with IR stability region $\xi > 0, \varepsilon < 0$;
\item FP8 $g^{*2}=0, y^*=-8\xi/3, \alpha^*=1$ with empty IR stability region. 
\end{itemize}

\begin{figure}[H]
	\centering
	\includegraphics[width=6 cm]{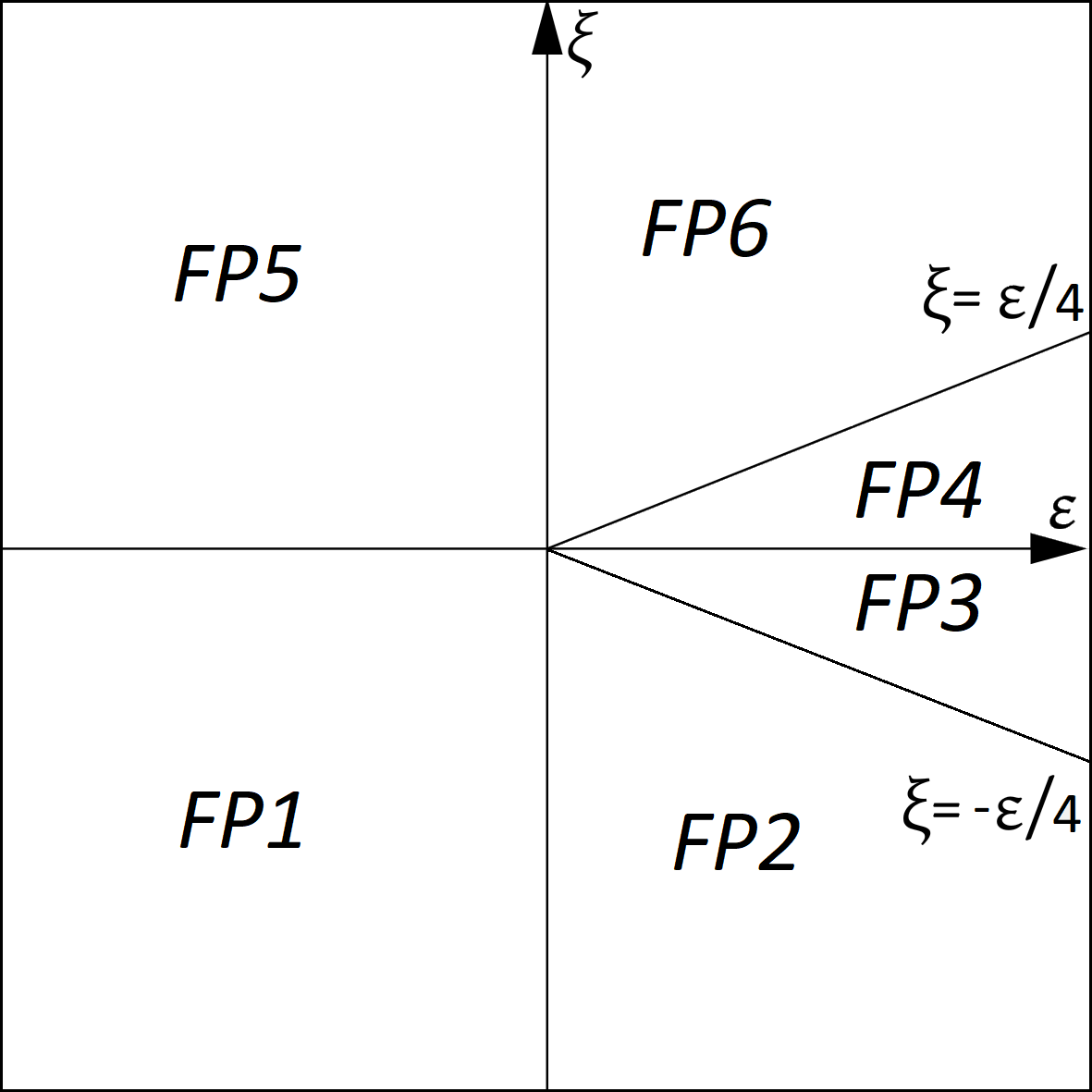}
	\caption{IR stability regions of the fixed points for the renormalized theory~(\ref{eq9}).
		\label{fig1}} 
\end{figure}

Stability regions on the $\varepsilon-\xi$ plane are shown at the Figure~\ref{fig1}.

Let us discuss the stability regions in details. The domain in the lower left part of the graph on the Fig. \ref{fig1} (quadrant III of $\varepsilon-\xi$ plane) corresponds to the couple of Gaussian points FP1a and FP1b (their combined stability regions are denoted as FP1 on the Fig. \ref{fig1}). This region is related to ordinary diffusion as all the coupling constants in the IR limit tend to zero for the negative $\xi$ and $\varepsilon$. 

The domain denoted FP2 corresponds to the regime where only the KPZ nonlinearity is relevant. As one can see, $x^*=0$ in this regime, which is consistent with the pure KPZ model (with the spatially quenched noise and without the turbulent mixing) being multiplicatively renormalizable without an extra term $h'{\bf v}^2$ in the action functional \cite{Stat}.

Surprisingly, while stability region of the point FP6 also corresponds to the regime where the KPZ nonlinearity is relevant and the turbulent mixing is irrelevant, the coordinate $x^*$ is not trivial for this regime: $x^*\neq 0$. That means that the new nonlinearity $h'{\bf v}^2$ is relevant even while the turbulent mixing (that induced it in the first place) is ``turned off''. It should be noted that it is this regime that corresponds to the Kolmogorov value of the exponent $\xi = 4/3$ (for either $d=2$ or $d=3$).

This is also the case with the regime for the point FP5: the new term $h'{\bf v}^2$ determines the IR asymptotic behaviour while both the turbulent mixing and the KPZ nonlinearity are irrelevant.

Finally, in the regimes corresponding to the points FP3 and FP4, both nonlinearities and the turbulent mixing are relevant as all coordinates of the fixed points are nontrivial.

It should be noted that for the regimes that correspond to the regions in the right-hand side of the graph on Fig. \ref{fig1} some coordinates of the fixed points are imaginary while some are negative.


\subsection{Critical exponents}


Note that the substitution $g \rightarrow g^*$ turns the RG equation~(\ref{eq13}) into an equation with constant coefficients
\begin{equation}
	\left(D_\mu-\gamma^{*}_\nu D_\nu -\gamma^{*}_{G_R}\right)G_R=0
	\label{eq17}
\end{equation}
Here $\gamma^{*}_\nu=\gamma_\nu (g^*, w^*,x^*)$ and $\gamma^{*}_{G_R}=\gamma_{G_R} (g^*, w^*,x^*)$.
Canonical scale invariance for the renormalized Green's function $G_R$ is expressed by the following differential equations:
\begin{equation}
	\begin{gathered}
		\left(\sum_i d_i^k D_i- d_{G_R}^k\right)G_R=0;  \\	
		\left(\sum_i d_i^\omega D_i- d_{G_R}^\omega\right) G^R=0.
	\end{gathered}
\label{eq18}
\end{equation}
Here $i$ a is full set of arguments of $G^R$: $\{i\}=\{\omega, k, g, w, x, \nu, \mu\}$. As before, $D_i=i\partial_i$ while $d^k$ and $d^\omega$ stand for canonical dimensions.
By combining~(\ref{eq17}) and~(\ref{eq18}) to eliminate operator $D_{\mu}$ (as $\mu$ is fixed in IR asymptotic) we obtain the equation of critical scaling:
\begin{equation}	
	\left(\Delta_\omega D_\omega+\Delta_k D_k-\Delta_{G}\right)G_R=0,
\end{equation}
where critical dimensions $\Delta_G$, $\Delta_k$, $\Delta_{\omega}$ are 
\begin{equation}
	\begin{gathered}
	\Delta_G=\gamma_{G_R}^{*}+d_{G_R}^k+d_{G_R}^\omega \Delta_{\omega}, \\ 
	\Delta_k=1, \quad
	\Delta_\omega=(2-\gamma_\nu^{*}). \\
	\end{gathered}
\label{eq20}
\end{equation}

Critical exponents $\chi$ and $z$ in the power law~(\ref{scaling}) are related to critical dimensions in a trivial way: $\chi=-\Delta_h$, $z=\Delta_{\omega}$. This statement, however, is not obvious and requires justification. Indeed, the power law~(\ref{scaling}) is written for the structure functions $S_n$ that consist of pair correlation functions $\langle h^s(x)h^q(0)\rangle$ of the composite fields $h^n(x)$ (``composite
operators''). Generally, renormalization of such objects requires additional (sometimes quite complex) analysis. In the present case, however, one can prove (see \cite{Us} for a detailed proof in a similar case) that operators $h^n$ are not renormalized and their critical dimensions are given by the equality $\Delta_{h^n}=n\Delta_h$. The latter is enough to justify the relation for critical exponents. 

Critical dimensions for the fixed points FP1--FP6 (one-loop approximation) are presented in the Table \ref{tab3}.

\begin{table}[H]
	\caption{Critical dimensions for the fixed points FP1--FP6. \label{tab3} }
	\begin{center}
	\begin{tabular}{ |c|c|c|c|c| }
		\hline
		$FP$&$\Delta_\omega$&$\Delta_h$&$\Delta_v$&$\Delta_{h'}$ \\ \hline
		
		$FP{1a}$, $FP{1b}$&$2$&$ d \slash 2 - 2$&$1-\xi \slash 2$&$d \slash 2 + 2$ \\ \hline
		
		$FP2$&$d \slash 4 +1$&$0$&$1 \slash 2-d \slash 8 - \xi \slash 2$&$d$ \\ \hline
		
		$FP3$ & $2+\xi$&$0$&$1$&$d$ \\ \hline
		
		$FP4$&$2-\xi$&$0$&$1-\xi$&$d$ \\ \hline
		
		$FP5$&$2$&$d \slash 2 -2$&$1-\xi \slash 2$&$d \slash 2 +2$ \\ \hline
		
		$FP6$&$d \slash 4 +1$&$0$&$1 \slash 2-d \slash 8-\xi \slash 2$&$d$ \\ \hline
	\end{tabular}
	\end{center}
\end{table}

Critical exponent $\Delta_h$ for the point FP2 (the regime of the pure KPZ model) is in agreement with \cite{Stat} (equation (43)) where the KPZ model with the spatially quenched noise was considered without the turbulent mixing. IR stability region of the point FP7 is empty but it is this point that corresponds to the regime of the pure turbulent mixing. Critical exponents for FP 7 related to the velocity field ($\Delta_{\omega}=2-\xi$, $\Delta_{v}=1-\xi$) are in agreement with exponents for Kazantsev–Kraichnan velocity ensemble, see, e.g. equations (2.18)--(2.20) in \cite{expo}. All critical exponents for the point FP4 coincide with the ones calculated for nontrivial regime of a system described by the original KPZ model (with the white random noise) and Kazantsev–Kraichnan ensemble, see equation (6.4) in \cite{Us}.
  

\section{Conclusion and Discussion}


We studied the KPZ model with a time-independent (columnar or spatially quenched) 
random noise and turbulent motion of the environment. The latter was simulated by the Kazantzev-Kraichnan's ``rapid-change''  velocity ensemble.
The problem was reformulated as a certain field-theoretic
model, and the standard renormalization procedure was applied. It was shown that the original model is not closed with respect to renormalization in the following sense: a new interaction (quadratic in the velocity field) unavoidably appears as a counterterm.

The general RG ideology requires this term (that was absent from the initial ``naive'' formulation of the model) to be included into consideration from the very beginning.
For the extended renormalizable field theoretic model, possible types of IR asymptotic behaviour (universality classes) are associated with the fixed points of the RG equations.

The RG equations, derived for our properly extended model with the new term, have eight fixed points. Two of them are always unstable,
while the others can be IR attractive for a certain choice of the spatial dimension $d$ and the velocity parameter $\xi$. Critical exponents that describe the IR  (long-time, large-distance) asymptotic behaviour of the correlation functions were  found in the leading order of the double expansion in $\varepsilon=4-d$ and $\xi$ (one-loop approximation). Their values are universal in the sense that they depend only on the spatial dimension $d$ and the parameter $\xi$.

All nontrivial scaling regimes correspond to complex or negative fixed points coordinates, which is a feature shared with the majority of models within the KPZ family. 

The most realistic values of parameters ($d=1,2,3$ and $\xi=4/3$) correspond to the point referred to as FP6 in Section \ref{FPs}. In this regime, the advection term appears IR irrelevant (in the sense of Wilson) while the new term is relevant along with the original KPZ nonlinearity.
This means that  the effect of the turbulent environment manifests itself not as a habitual transfer (advection) but as a certain nonlinear interaction of the velocity field. 

This situation has some interesting parallels  with the light-light (or photon-photon) scattering, the phenomenon that is absent in the classical electrodynamics but emerges in the quantum
case as a result of interaction with the vacuum fluctuations. To be precise, the situation can roughly be compared with the weak-field limit of the Euler-Heisenberg electrodynamics that involve a term, quartic in the electromagnetic potential \cite{EH}. 
The simplest interaction term responsible for such phenomenon is a local term quartic in the electromagnetic potential, $(A_{\mu}A^{\mu})^2$. It has the necessary canonical dimension but it is forbidden by the gauge symmetry and, therefore, cannot emerge as a counterterm. Similarly, dimensional considerations show that the term ${\bf v}^2$ can be added to the KPZ equation, no matter what kind of random noise is used. However, if we require the model to be Galilean covariant, such term is forbidden. This happens in the case of the white in-time noise because it preserves Galilean symmetry. On the contrary, in our model, this symmetry is already violated by the spatially quenched noise so the term $h'{\bf v}^2$ necessarily results from the renormalization procedure and should be added into the action functional from the very beginning. 

However, quartic terms in QED do appear in the effective action functional due to the radiative corrections. For the first time, they were derived within the Euler-Heisenberg Lagrangian and, for small fields, have the forms $({\bf E}^2-{\bf B}^2)^2$ and $({\bf EB})^2$, which are Lorentz and gauge invariant.
Written in terms of the potential $A_{\mu}$, they involve fourth-order derivatives and have larger canonical dimension in comparison with the classical Lagrangian $L \sim ({\bf E}^2-{\bf B}^2)$. This difference is compensated by the dimensional coefficient $1/m_e^4$, where $m_e$ is the electron mass.

Similarly, the effective action for the KPZ model in the Galilean invariant case may include invariant terms, quadratic in the gradients of the velocity field (resulting from the 1-irreducible function $\langle h'{\bf vv}\rangle$) with larger canonical dimension and the dimensional coefficient $1/m^2$.

For the heat transfer equation, an explicit expression for those terms is presented in \cite{Landau}, see equation (50.2) on page 197. They describe dissipation of the mechanical kinetic energy of the fluid into heat and lead to the linear growth of the mean temperature.
Note that in the energy balance equation for the fluid they appear with the opposite sign; see e.g. \cite{RedBook}. 

It should be stressed that the actual dimensionless parameter in the gradient expansion for the KPZ equation is $\partial^2/m^2 \sim k^2/m^2 \ll 1$. In the Galilean invariant case (white in-time noise)
the contribution of the new terms to the surface growth rate is suppressed by this parameter, while in the present case (spatially quenched noise) the effect is of order $O(1)$ and, therefore, is much more strongly pronounced. Thus, the RG ideology requires that this $h'{\bf v}^2$ term be included in the action functional from the very start. 

The new term $h'{\bf v}^2$ also determines the IR asymptotic behaviour in the regime that corresponds to the fixed point FP5, where both the turbulent mixing and the KPZ nonlinearity are irrelevant.

Now  let us briefly comment on the physical interpretation of the negative and imaginary fixed points coordinates. Imaginary values of the coordinate $g^*$ correspond to a negative amplitude of the pair correlator for the field $h$. This has several possible implications. Firstly, it may imply a connection to models constructed with the Doi-Peliti formalism \cite{Doi,Tauber} where quadratic terms with negative signs and imaginary random noise can appear \cite{Tauber,Benitez1}. Secondly, there is a mapping of the KPZ equation with the white in-time noise onto the one-dimensional Lieb--Liniger model of Bose gas \cite{Bose,Bose2}. Surprisingly, it is negative sign of the pair correlator that corresponds to the Bose gas with repulsion \cite{BDC}. Lastly, such coordinates do not preclude the existence of non-perturbative IR attractive fixed point that remains hidden in perturbative RG analysis. See a more detailed discussion of related issues in \cite{AKL}.

Elsewhere, negative values of kinetic coefficients were sometimes encountered in non-equilibrium stochastic models, especially involving compressible fluids  \cite{Yakhot,AvelVerg}.
Complex effective viscosity coefficient is featured in stochastic equations for Langmuir plasma turbulence \cite{plasma,plasma1} and in a stochastic version of the nonlinear Schr\"{o}dinger equation \cite{Tauber2}. Imaginary fixed points 
and negative contributions to the diffusivity coefficient
were recently obtained in a model of active scalar turbulent convection \cite{Zapiski}.

Thus, the problem of complex effective values of real physical quantities 
appears ubiquitous and pervasive in non-equilibrium stochastic problems. 
The physical interpretation of this persistent phenomenon deserves a careful and systematic analysis and suggests  important directions in further investigation.
In particular, it is interesting to study more realistic and complex models. 
As possible generalizations of our present model, non-Gaussian velocity fields with finite correlation time, governed by various types of stochastic Navier-Stokes equations, can be employed. This work is already in progress.


\section*{Acknowledgments}


The work of P.I.K., N.M.G. and N.V.A. was funded by the Russian Foundation for Basic Research~(RFBR), project number~20-32-70139. 
The work of M.M.T. was supported by the RFBR, project number~19-32-60065.


\reftitle{References}


\begin{thebibliography}{999}

\bibitem{KPZ} 
Kardar M.; Parisi G.; Zhang Y-C. Dynamic Scaling of Growing Interfaces {\it Phys. Rev. Lett.} {\bf 1986}, {\it 56}, 889

\bibitem{rost1} 
Krug J.; Spohn H. {\it Solids far from equilibrium}, Ed. Godreche C Cambridge: Cambridge University Press, 1990. 

\bibitem{rost2} 
Halpin-Healy T.; Zhang Y-C. Kinetic roughening phenomena, stochastic growth, directed polymers and all that. Aspects of multidisciplinary statistical mechanics {\it Phys. Rep.} {\bf 1995} {\it 254} 215 

\bibitem{FNS1} Forster D.; Nelson D. R.; Stephen M. J. Long-Time Tails and the Large-Eddy Behavior of a Randomly Stirred Fluid, 
{\it Phys. Rev. Lett.} {\bf 1976}, {\it 36}, 867.

\bibitem{Nobel} Giorgio Parisi -- Facts -- 2021. NobelPrize.org. Nobel Prize Outreach AB 2021. Mon. 11 Oct 2021. https://www.nobelprize.org/prizes/physics/2021/parisi/facts/

\bibitem{Mitch}  Feigenbaum  M.J.; Procaccia~I.; Davidovich~B.  Dynamics of Finger Formation in Laplacian Growth Without Surface Tension
{\it J.Stat. Phys.} {\bf 2001}, {\it 103}, 973–1007

\bibitem{Zia} Schmittmann B.; Zia R.K.P. Driven diffusive systems: An introduction and recent developments {\it Phys. Rep.} {\bf 1998}, {\it 301}, Issues 1–3, p. 5-64.

\bibitem{UFN}
 Adzhemyan L.Ts.;  Antonov N.V.;  Vasil’ev A.N. Quantum field renormalization group in the theory of fully developed turbulence {\it Usp. Fiz. Nauk} {\bf 1998}, {\it 166}, 1257 [In Russian, Engl. Transl.: {\it Phys.–Usp.} {\bf 1996}, {\it 39}, 1193.

\bibitem{RedBook} 
Adzhemyan L.Ts.;  Antonov N.V.;  Vasil’ev A.N. 
{\it The Field Theoretic Renormalization Group in Fully Developed Turbulence} Gordon \& Breach, London, 1999.

\bibitem{Bak} Bak P. {\it How Nature Works: The Science of Self-Organized Criticality}, Copernicus, N.Y., 1996.

\bibitem{Bak3} 
G. Pruessner {\it Self-Organized Criticality: Theory, Models and
Characterisation} Cambridge University Press: Cambridge, MA, USA, 2012.

\bibitem{Col1} 
 Mu\~{n}oz M.A. Colloquium: Criticality and dynamical scaling in living systems {\it Rev. Mod. Phys.} {\bf 2018}, {\it 90} 031001.

\bibitem{Col2} 
Markovi\'c D.; Gros C. Power laws and self-organized criticality in theory and nature {\it Phys. Rep.} {\bf 2014}, {\it 536}, 41.

\bibitem{Corwin} Corwin I. The Kardar-Parisi-Zhang equation and universality class {\it Random Matrices: Theory and Applications} {\bf 2012}, {\it 1(1)}, 1130001.

\bibitem{freshrev} 
Takeuchi K. A. An appetizer to modern developments on the Kardar–Parisi–Zhang universality class. {\it Physica} A {\bf 2018}, {\it 504}, 77

\bibitem{ff} 
Strack P. Dynamic criticality far from equilibrium: One-loop flow of Burgers-Kardar-Parisi-Zhang systems with broken Galilean invariance. {\it  Phys. Rev.} E {\bf 2015}, {\it 91}, 032131 

\bibitem{Sinc}  
Niggemann O.;  Hinrichsen H. Sinc noise for the Kardar-Parisi-Zhang equation. {\it Phys. Rev.} E {\bf 2018} {\it 97}, 062125 

\bibitem{Uppergt41}  
Katzav E.;  Schwartz M. Existence of the upper critical dimension of the Kardar–Parisi–Zhang equation. {\it Physica} A {\bf 2002}, {\it 309}, 69 

\bibitem{UpperInf}
Alves S.G.;  Oliveira T.J.;  Ferreira S.C. Universality of fluctuations in the Kardar-Parisi-Zhang class in high dimensions and its upper critical dimension. {\it  Phys. Rev.} E,  {\bf 2014}, {\it 90}, 020103(R) 

\bibitem{Polaritons} Altman E.; Sieberer L. M.; Chen L.; Diehl S.; Toner J. Two-dimensional superfluidity of exciton polaritons requires strong anisotropy. {\it Phys. Rev.} X, {\bf 2015}, {\it 5}, 011017.

\bibitem{Polaritons1} 
Ji K., Gladilin V. N.; Wouters M. Temporal coherence of one-dimensional nonequilibrium quantum fluids. {\it Phys. Rev.} B, {\bf 2015}, {\it 91} 045301.

\bibitem{Polaritons2} Deligiannis K.; Squizzato D.; Minguzzi A.; Canet L. Accessing Kardar-Parisi-Zhang universality sub-classes with exciton polaritons (a). {\it EPL} {\bf 2020}, {\it 132}, 67004.

\bibitem{CanetColor} Squizzato D.; Canet L. Kardar-Parisi-Zhang equation with temporally correlated noise: A nonperturbative renormalization group approach {\it Phys. Rev.} E,  {\bf 2020}, {\it 100}, 062143.

\bibitem{Sky} Najem S.; Krayem A.; Ala-Nissila T.; Grant M. Kinetic roughening of the urban skyline. {\it Phys. Rev.} E, {\bf 2020}, {\it 101}, 050301(R).

\bibitem{11} 
L\"{a}ssig M. On the renormalization of the Kardar-Parisi-Zhang equation. {\it Nucl. Phys.} B, {\bf 1995}, {\it 448}, 559.

\bibitem{111}
Wiese K.J. On the perturbation expansion of the KPZ equation {\it J. Stat. Phys.} {\bf 1998}, {\it 93}, 143. 

\bibitem{Canet} 
Canet L.; Chat\'{e} H.;  Delamotte B.;  Wschebor N. Nonperturbative renormalization group for the Kardar-Parisi-Zhang equation  {\it Phys. Rev. Lett.} {\bf 2010}, {\it 104}, 150601. 

\bibitem{Canet1}  
Canet L.; Chat\'{e} H.;  Delamotte B.;  Wschebor N. Nonperturbative renormalization group for the Kardar-Parisi-Zhang equation: General framework and first applications. {\it Phys. Rev.} E,  {\bf 2011}, {\it 84}, 061128.

\bibitem{Canet2} 
Kloss T.; Canet L.; Wschebor N. Nonperturbative renormalization group for the stationary Kardar-Parisi-Zhang equation: Scaling functions and amplitude ratios in 1+1, 2+1, and 3+1 dimensions. {\it Phys. Rev.} E {\bf 2012}, {\it 86}, 051124.

\bibitem{Canet4}  
Mathey S.;  Agoritsas E.;  Kloss T.;  Lecomte V.;  Canet L. Kardar-Parisi-Zhang equation with short-range correlated noise: Emergent symmetries and nonuniversal observables. {\it Phys. Rev.} E {\bf  2017}, {\it 95}, 032117. 

\bibitem{LK} 
L\"assig M.;  Kinzelbach H. Upper critical dimension of the Kardar-Parisi-Zhang equation. {\it Phys. Rev. Lett.} {\bf 1997}, {\it 78}, 903. 

\bibitem{LK1}
Colaiori F.;  Moore M. Upper critical dimension, dynamic exponent, and scaling functions in the mode-coupling theory for the Kardar-Parisi-Zhang equation. {\it Phys. Rev. Lett.} {\bf 2001}, {\it 86}, 3946. 

\bibitem{UpperInf0}
Marinari E.;  Pagnani A.;  Parisi G.;  Ra\'cz Z. Width distributions and the upper critical dimension of Kardar-Parisi-Zhang interfaces. {\it Phys. Rev.} E {\bf  2002}, {\it 65}, 026136. 

\bibitem{Upper4}  
Fogedby H.C. Localized growth modes, dynamic textures, and upper critical dimension for the Kardar-Parisi-Zhang Equation in the weak-noise limit. {\it Phys. Rev. Lett.} {\bf 2005}, {\it 94} 195702;\\
Kardar-Parisi-Zhang equation in the weak noise limit: Pattern formation and upper critical dimension. {\it Phys. Rev.} E {\bf 2006}, {\it 73}, 031104; \\ 
Patterns in the Kardar-Parisi-Zhang equation. {\it J. Phys. (Pramana)} {\bf 2008}, {\it 71}, 253. 

\bibitem{Uppergt4}  
Katzav E.;  Schwartz M. Existence of the upper critical dimension of the Kardar–Parisi–Zhang equation. {\it Physica} A {\bf 2002}, {\it 309}, 69. 

\bibitem{Howard}  
T\"auber U.C.;  Howard M.;  Vollmayr-Lee B.P. Applications of field-theoretic renormalization group methods to reaction–diffusion problems. {\it J. Phys. A: Math. Gen.} {\bf 2005}, {\it 38},  R79. 

\bibitem{Cooper} 
Cooper F.; Dawson J.F. Auxiliary field loop expansion of the effective action for a class of stochastic partial differential equations {\it Annals of Physics} {\bf 2016}, {\it 365}, 118. 

\bibitem{Pav} 
Pavlik S.I. Scaling for a growing phase boundary with nonlinear diffusion. {\it JETP} {\bf 1994}, {\it 79}, 303 
[Translated from the Russian: {\it ZhETF} {\bf 1994}, {\it 106}, 553]
 
\bibitem{Pav2}
Antonov N.V.;  Vasil'ev A.N. The quantum-field renormalization group in the problem of a growing phase boundary {\it JETP} {\bf 1995}, {\it 81}, 485 [Translated from the Russian: {\it ZhETF} {\it 108} 885]

\bibitem{Caldarelli} Caldarelli G.; Giacometti A.; Maritan A.; Rodriguez-Iturbe I.; Rinaldo A. Randomly pinned landscape evolution.  {\it Phys. Rev.} E {\bf 1997}, {\it 55}, R4865(R).

\bibitem{Czi} Czir\'{o}k A.; Somfai E.; Vicsek J. Experimental evidence for self-affine roughening in a micromodel of geomorphological evolution. {\it Phys. Rev. Lett.} {\bf 1993}, {\it 71}, 2154.

\bibitem{JP} Lee C.;  Kim J.M. Depinning transition of the quenched Kardar-Parisi-Zhang equation. {\it Journal of the Korean Physical Society}, {\bf 2005}, {\it 47(1)}, 13. 

\bibitem{KimKim} Jeong H.; Kahng B.; Kim D. Anisotropic surface growth model in disordered media.
{\it Phys. Rev. Lett.} {\bf 1996}, {\it 25}, 5094.

\bibitem{KimKimKim}
 Kim H.-J.;  Kim I.-m.; Kim J. M. Hybridized discrete model for the anisotropic Kardar-Parisi-Zhang equation.
{\it Phys. Rev.} E {\bf 1998}, {\it 58}, 1144.

\bibitem{Hinri} Hinrichsen H. Non-equilibrium critical phenomena and phase transitions into absorbing states.  {\it Adv. Phys.} {\bf 2000}, {\it 49}, 815-958.

\bibitem{Nara} Narayan O.; Fisher D.S. Threshold critical dynamics of driven interfaces in random media.
{\it Phys, Rev.} B {\bf 1993}, {\it 48}, 7030.

\bibitem{Janssen}  Janssen H.K. Renormalized field theory of the Gribov process with quenched disorder {\it Phys, Rev.} E {\bf 1997}, {\it 55}, 6253.

\bibitem{Moreira}  Moreira A.G.; Dickman R.  Critical dynamics of the contact process with quenched disorder. {\it Phys. Rev.} E {\bf 1996}, {\it 54}, R3090.

\bibitem{Webman} Webman I.; ben Avraham D.; Cohen A.; Havlin S. Dynamical phase transitions in a random environment  {\it Phil. Mag.} B {\bf 1998}, {\it 77}, 1401.

\bibitem{Delamotte} Duclut C.; Delamotte B. Nonuniversality in the erosion of tilted landscapes. {\it Phys. Rev.} E {\bf 2017}, {\it 96}, 012149.

\bibitem{Vitalik} Antonov N.V.; Gulitskiy N.M.;  Kakin P.I.; Serov V.D. Effects of turbulent environment and random noise on self-organized critical behavior: Universality versus nonuniversality.
{\it Phys. Rev.} E {\bf 2021}, {\it 103}, 042106.

\bibitem{CKPZSQN} Mukherjee S. Conserved Kardar-Parisi-Zhang equation: Role of quenched disorder in determining universality. {\it Phys. Rev.} E {\bf 2021}, {\it 103}, 042102.

\bibitem{Onuki2}
Imaeda T,; Onuki A.; Kawasaki K. Anisotropic spinodal decomposition under shear flow {\it Progr. Theor. Phys.} {\bf 1984}, {\it 71}, 16.

\bibitem{Satten2} 
Satten G.; Ronis D. Critical phenomena in randomly stirred fluids: Correlation functions, equation of motion, and crossover behavior. {\it Phys. Rev.} A {\bf 1986}, {\it 33}, 3415.

\bibitem{Nelson}
Aronowitz A.; Nelson D.R. Turbulence in phase-separating binary mixtures. {\it Phys. Rev.} A {\bf 1984}, {\it 29}, 2012.

\bibitem{AHH} Antonov N.V.; Hnatich M.;  Honkonen J. Effects of mixing and stirring on the critical behaviour.
{\it J. Phys. A: Math. Gen.} {\bf 2006}, {\it  39}, 7867.

\bibitem{Alexa} Antonov N.V.; and  Ignatieva A.A. Critical behaviour of a fluid in a random shear flow: Renormalization group analysis of a simplified model
{\it J. Phys. A: Math. Gen.} {\bf 2006}, {\it 39}, 13593.

\bibitem{AIK}
Antonov N.V.; Iglovikov V.I.;  Kapustin A.S. Effects of turbulent mixing on the nonequilibrium critical behaviour.
{\it J. Phys. A: Math. Theor.} {\bf 2008}, {\it 42}, 135001.

\bibitem{AKM} Antonov N.V.;  Kapustin A.S.; Malyshev A.V. Effects of turbulent transfer on critical behavior {\it Theor. Math. Phys.} {\bf 2011}, {\it 169} 1470.

\bibitem{Us} 
Antonov N.V.; Kakin P.I. Random interface growth in a random environment: Renormalization group analysis of a simple model. {\it  Theor. Math. Phys.} {\bf 2015}, {\it 185}, 1391. 

\bibitem{AKL} 
Antonov N.V.; Kakin P.I.; Lebedev N.M. The Kardar–Parisi–Zhang model of a random kinetic growth: effects of a randomly moving medium.  {\it J. Phys. A: Math. Theor.} {\bf 2019}, {\it 52}, 505002.

\bibitem{Script} Antonov N.V.; Gulitskiy N.M.; Kakin P.I.; Kostenko M.M. Effects of turbulent environment on the surface roughening: The Kardar-Parisi-Zhang model coupled to the stochastic Navier–Stokes equation. {\it Phys. Scr.} {\bf 2020}, {\it 95}, 084009.

\bibitem{FGV} 
Falkovich G.; Gaw\c{e}dzki K.; Vergassola M. Particles and fields in fluid turbulence. {\it Rev. Mod. Phys.} {\bf 2001}, {\it 73}, 913.

\bibitem{Vasiliev} 
Vasiliev A.N. {\it The Field Theoretic Renormalization Group in Critical behaviour Theory and Stochastic Dynamics} (Chapman \& Hall/CRC, Boca Raton) 
[Translated from the Russian: 1998 (St Petersburg, Institute of Nuclear Physics, Gatchina, ISBN 5-86763-122-2)]

\bibitem{Stat} Antonov N.V.; Kakin P.I.; Lebedev N.M. Static Approach to Renormalization Group Analysis of Stochastic Models with Spatially Quenched Noise. {\it J.~Stat. Phys.} {\bf 2020}, {\it 178}, 392.

\bibitem{expo} Adzhemyan L.T.; Antonov N.V.; Vasiliev A.N. Renormalization group, operator product expansion, and anomalous scaling in a model of advected passive scalar. {\it Phys. Rev.} E {\bf 1998}, {\it 58}, 1823.

\bibitem{EH} Heisenberg W.; Euler H. Folgerungen aus der Diracschen Theorie des Positrons {\it Z.Phys.} {\bf 1936}, {\it 98}, p. 714-732.

\bibitem{Landau} Landau L.D.; Lifshitz E.M. {\it  Fluid Mechanics} (Vol.6 of Course of Theoretical Physics, 2nd English edition) Pergamon Press, Oxford, 1987

\bibitem{Doi} Doi M. Stochastic theory of diffusion-controlled reaction. {\it J. Phys.} A {\bf 1976}, {\it 9}, 1479; \\
Grassberger P.; Scheunert P. Fock‐space methods for identical classical objects. {\it Fortschr. Phys.} {\bf 1980}, {\it 28}, 547; \\
L.~Peliti, {\it J. Phys.} (Paris) {\bf 1984}, {\it 46}, 1469.

\bibitem{Tauber} T\"{a}uber U.C., Dynamic Phase Transitions in Diffusion-Limited Reactions. {\it Acta Physica Slovaca} {\bf 2002}, {\it 52}, 505;  \\ T\"{a}uber U.C. Scale invariance and dynamic phase transitions in diffusion-limited reactions {\it Adv. Solid State Phys.} {\bf 2003}, {\it 43}, 659; \\
T\"{a}uber U.C., {\it Lect. Notes Phys.} {\bf 2007}, {\it 716}, 295; \\ 
T\"{a}uber U.C., In: {\it Encyclopedia of Complexity and System Science}
R.A.~Meyers (ed.) (Springer, New York) 3360 (2009)

\bibitem{Benitez1} Benitez F.; DuclutC.; Chaté H.; Delamotte B.; Dornic I.; Muñoz M. A. Langevin equations for reaction-diffusion processes. {\it Phys. Rev. Lett.} {\bf 2016}, {\it 117}, 100601.

\bibitem{Bose} Roberts J.L.; Claussen N.R.; Cornish S.L.; Donley E.A.; Cornell E.A.; Wieman C.E. Controlled collapse of a Bose-Einstein condensate. {\it  Phys. Rev. Lett.} {\bf 2001}, {\it 86}, 4211.

\bibitem{Bose2}  Lieb E.H.;  Liniger W. Exact analysis of an interacting Bose gas. I. The general solution and the ground state. {\it Phys. Rev.} {\bf 1963}, {\it 130(4) }, 1605.

\bibitem{BDC} Busiello G.; De Cesare L. The critical exponents $\eta$ and $z$ to second order in $\epsilon = 2-d$ for a Bose system at $T= 0$ {\it Phys. Lett.} {\bf 1980}, {\it 77}A, 177.
 
\bibitem{Yakhot}
Yakhot V.
Ultraviolet dynamic renormalization group: Small-scale properties of a randomly stirred fluid,
{\it Phys. Rev.} A {\bf 1981}, {\it 23},1486-1497; \\
Yakhot V. Large-scale properties of unstable systems governed by the Kuramoto-Sivashinksi equation,
{\it Phys. Rev.} A {\bf 1981}, {\it 24 }, 642(R)-644(R); \\
Sivashinsky G.; Yakhot V., 
Negative viscosity effect in large-scale flows, {\it Physics of Fluids} {\bf 1985}, {\it 28}, 1040-1042.

\bibitem{AvelVerg} Avellaneda M.; Vergassola M. Scalar transport in compressible flow. {\it Physica D: Nonlinear Phenomena}
{\bf 1997}, {\it 106}, 148-166. 

\bibitem{plasma} Pelletier G. Langmuir turbulence as a critical phenomenon. Part 2. Application of the dynamical renormalization group method. {\it J. Plasma Phys.} {\bf 1980}, {\it 24}, 421-443. 

\bibitem{plasma1}  Adzhemyan L.Ts.;  Vasil'ev A.N.;  Gnatich M.;  Pis'mak Yu.M.
Quantum field renormalization group in the theory of stochastic Langmuir turbulence.
{\it Theor. Math. Phys.} {\bf 1989}, {\it 78 }, 260–272.

\bibitem{Tauber2} T\"{a}uber U.C.;  Diehl S. Perturbative field-theoretical renormalization group approach to driven-dissipative Bose-Einstein criticality. 
{\it Phys. Rev.} X {\bf 2014}, {\it 4 },  021010.

\bibitem{Zapiski} Antonov N.V.; Kostenko M.M. Renormalization Group in the Problem of Active Scalar Advection.  
{\it J. Math. Sci.} {\bf 2021}, {\it 257}, 425-441.
 

\end{thebibliography}
\end{document}